\documentclass[12pt]{article}

\title{Phenomenology of  ``dark matter''---from the Everett's quantum cosmology}
\author{Michael B. Mensky\\
{\small P.N.Lebedev Physical Institute,} 
{\small 53 Leninsky prosp., 119991 Moscow, Russia}}
\date{18th of May, 2011}

\begin{document}

\maketitle

\begin{abstract}
It is widely accepted that the Everett's (or ``many-worlds'') interpretation of quantum mechanics is the only one which is appropriate for quantum cosmology because no environment may exist for Universe as a whole. We discuss, in the framework of the Everett's interpretation, the (quasi-) classical stage of evolution of the Universe when there coexist ``classically incompatible'' configurations of matter, or classical alternative realities (``alternatives'' for short). In the framework of the Everett's interpretation the semiclassical gravity (where the gravitational field is classical and the non-gravitational fields are quantum) is more natural than theories including quantizing gravitational field. It is shown that the semiclassical (at least on the astrophysical and cosmological scales) Everett-type gravity leads to the observational effect known as the effect of dark matter. Instead of assuming special forms of matter (weakly interacting with the known matter), the role of the dark matter is played in this case by the matter of the usual kind which however belongs to those alternative realities (Everett's worlds) which remain ``invisible'', i.e. not perceived with the help of non-gravitational fields.
\end{abstract}


\newpage

\section{Introduction}
\label{sec:Intro}

Quantum cosmology possesses essential conceptual distinctions as compared to quantum theory of microscopic objects. One of them is connected with the fact that Universe as a whole includes everything that exists. Therefore the standard Copenhagen interpretation of quantum mechanics, suggesting a measuring device (or measuring environment) existing outside the system under consideration, is not in fact applicable in quantum cosmology.  

This is why some authors develop quantum cosmology in the framework of the Everett's (or ``many-worlds'') interpretation of quantum mechanics \cite{Everett} in which macroscopically distinct states of the world (Everett's worlds, or alternative classical realities, or alternatives for short) can coexist. Various ways to apply the Everett's theory in quantum cosmology were proposed, see for example \cite{Hartle2005, Halliwell1992}. 

Coexisting ``classical alternatives'' in the form of components of a superposition is assumed in every work of this type. Our proposal differs in the way of accounting the gravitational action of these alternatives on each other. 

The hypothesis will be accepted that the gravitational field, which acts on the matter in any alternative, is common for all alternatives. Technically this common gravitational field is determined by \emph{the energy-momentum tensor averaged over all alternatives}, as in the semiclassical approximation for quantum gravity. We shall see that this assumption has essential \emph{conceptual advantages in describing localization} of matter: it allows one to correctly say of the pieces of matter \emph{located in the same place} but belonging to different alternatives. 

Because of these features of theory, astrophysical observations denoted as \emph{the effects of dark matter} may be explained without postulating unknown forms of matter. The principle of the explanation is following. 

In the framework of the Everett's interpretation, the alternative classical realities are separated in consciousness. This means that \emph{subjectively} only a single alternative is perceived, while all the rest remain ``invisible''. In our version of Everett's cosmology the separation of the alternative configurations of matter in the Universe takes place in respect of \emph{observations with the help of the non-gravitational fields}. If the matter of a certain alternative is visible in these fields (can be observed with the help of them), then the matter belonging to the rest alternatives is invisible. However, \emph{the matter of ``other alternatives'' reveals itself by its gravitational field}. This is just what is meant by the observational effect of dark matter. 

\newpage
\section{Everett's quantum cosmology}
\label{sec:EverettSQuantumCosmology}

\paragraph{Cosmic alternatives form Alterverse.}

It is often assumed in the works on quantum cosmology that our physical world has a complicated structure, consisting of multiple ``partial'' universes which together comprise everything that exists. Such a conglomerate of the physically existing partial universes is called ``Multiverse''. The visual image for a Multiverse is a number of bubbles, some of them appearing on the surface of others and each gradually inflating. 

Sometimes the term ``Multiverse'' is applied to the set of the ``Everett's worlds'', the concept characteristic for the Everett's (many-worlds) interpretation of quantum mechanics. However, this terminology is misleading and in fact incorrect. ``Everett's worlds'' are not different physical (material) worlds. The term ``Everett's worlds'', or more adequate term ``alternative classical realities'', denotes the classical counterparts, together consisting a quantum state of a single physical, or material, world. 

The situation is well illustrated by simple quantum formulas. Let our (quantum) world may be in one of the states $\Psi_i$ (with $i$ taking some set of values). Then, according to principles of quantum mechanics, our world may also be in the state $\Psi = \sum_i \Psi_i$. The actual state of the world $\Psi$ is then a superposition, or sum, of the states $\Psi_i$. According to Everett's interpretation each of the states $\Psi_i$ may be essentially classical (quasiclassical) but describing macroscopically distinct pictures of the world. The actual state $\Psi$ of the world may be characterized in this case only by the whole set of the partial states $\Psi_i$. However, all these partial states should be considered on equal foot, or as being ``equally real'', or ``coexisting'' (as the components of the superposition $\Psi$). 

One can say that the classical ``Everett's worlds'' $\Psi_i$ are only ``classical projection'' of the only objectively existing state $\Psi$ of the physical world. Other wording may be that $\Psi_i$ are ``alternative classical realities'', or simply alternatives, while the quantum reality is presented by $\Psi$, i.e. by the whole totality of the alternatives.  

The author suggested a special term for this situation to differ it from the situation of Multiverse. The set of all Everett's worlds (or alternative classical realities) forming a single state of the (quantum) physical world, may be called ``Alterverse'' (see \cite{TermAterverse, MBMbk2010} where this term has been used). 

In the context of quantum cosmology, the counterparts of Alterverse, i.e. the alternative classical realities, or Everett's worlds, may be called \emph{``cosmic alternatives''}. In the present paper cosmic alternatives represent alternative variants of the large-scale structure of Universe arising due to its quantum nature.

\paragraph{Cosmic alternatives are born at the borderline between quantum and classical regimes of the universe.}

Cosmic alternatives are born at the stage of evolution of Universe when the deep quantum regime of this evolution converts into the ``classical'', or rather quasiclassical regime of evolution. In the early (quantum) stage of the universe (period of its inflation), the matter which fills Universe, is a quantum scalar field called \emph{inflanton field} because the properties of this field originate the inflation of the universe. 

At the end of inflation, the character of evolution of Universe changes. Usually it is claimed that further evolution may be (with a good approximation) described as classical (not quantum), but the evolution may follow various (alternative) scenarios forming the set of cosmic alternatives. In the spirit of the Everett's interpretation we have to say that the post-inflation evolution is described as a superposition of all cosmic alternatives. 

Precise description of the transition from quantum to classical stage of the evolution of the universe is a complicated physical and mathematical problem, see for example \cite{Halliwell1992, Hartle2005}. In these and in many other works the problem pursued by the authors is to find the possible scenarios of evolution. This task is somewhat simpler if the alternative scenarios, or \emph{cosmic alternatives}, do not interfere (or almost do not interfere) and therefore may be considered independently from each other. 

We shall leave this issue aside and concentrate on the other. Our task will be to find the consequences of the fact that the alternatives are superposed, i.e. in a sense coexist. We are interested in the consideration not of single alternatives but of \emph{the whole set of these alternatives as components of a quantum superposition} (see Sect.~\ref{sec:Intro}). 

\newpage
\section{Semiclassical Gravity}
\label{sec:SemiclassicalGravity}

\paragraph{Gravitational field is common for all alternatives.}

The second very important difference of our consideration from the conventional way of consideration concerns the gravitational field created by the matter in the alternatives and acting on this matter. 

Usually it is assumed that the gravitational field in each classical alternative is created by the matter which is in the state characteristic for this alternative. We shall accept a hypothesis according to which \emph{the gravitational field is the same for all cosmic alternatives and is created by the state of the matter, averaged over all alternatives}. This common gravitational field satisfies the Einstein equation in the form
\begin{equation}\label{EinsteinSemiClassic}
	{\cal G}_{\mu\nu} = \frac{8\pi G}{c^4}\left\langle T_{\mu\nu}\right\rangle
\end{equation}
The left-hand-side here is the Einstein tensor of the gravitational field (common for all alternatives), while the energy-momentum tensor in the right-hand-side is averaged over the quantum state of the universe $\Psi=\sum_i \Psi_i$ (therefore, over all cosmic alternatives $\Psi_i$, forming this state). 

This form of Einstein equation is usually treated as so-called semiclassical approximation in quantum gravity. Some authors argue though \cite{SemiclassGravity2011} that gravitational field is fundamentally classical. The question is not finally solved, but up to now it is found no logical necessity or experimental evidence demanding for the gravitational field to be quantized. For our goal, it is essential that gravitational field might be considered classical at astrophysical scales. 

\paragraph{Non-conservation of visible matter.}

We shall see in Appendix~\ref{sec:PathGroupLocalization} that, due to this hypotheses, it is possible to locally compare the cosmic alternatives with each other, i.e. to compare different alternatives in the same space-time region. The reason is that \emph{the concept of localization becomes common for all alternatives}. As a result, we can compare the states of matter in different alternatives but in the same place. The great advantage of this feature of semiclassical gravity is that it allows one to explain the phenomenon discovered in astrophysical observations and called  ``the effect of dark matter''. 

One of the consequences (the so-called Bianki identity) of Eq.~(\ref{EinsteinSemiClassic}) is that tensor ${\cal T}_{\mu\nu} = \left\langle T_{\mu\nu}\right\rangle$ has null divergency: 
\begin{equation}\label{BiankiAveraged}
	D^\mu {\cal T}_{\mu\nu} = 0 
\end{equation}
where covariant derivative $D_\mu$ is due to the common gravitational field of all alternatives. This equation, which expresses covariant energy-momentum conservation law, is compatible with the analogous equation for the energy-momentum tensor of the matter in each alternative: 
\begin{equation}\label{BiankiAlternative}
	D^\mu T^{(i)}_{\mu\nu} = 0. 
\end{equation}
However, the Eq.~(\ref{BiankiAlternative}) does not follow from Eq.~(\ref{BiankiAveraged}). What does this means?

In the limit of very long time, when all alternatives become identical (see the end of Sect.~\ref{sec:DarkMatter}), the alternative energy-momentum tensors become equal to the averaged one, $T^{(i)}_{\mu\nu}={\cal T}_{\mu\nu}$. Eqs.~(\ref{BiankiAveraged}),~(\ref{BiankiAlternative}) become then equivalent. Therefore, at the time infinity the covariant energy-momentum conservation law becomes valid for each alternative separately. However, it is not valid for finite times. 

This means that energy-momentum can flow from one to the other alternatives under the condition that the complete energy-momentum is conserved. Energy belonging to a single alternative (say, one which is subjectively perceived) is not necessarily conserved. This may look as creation (or disappearance) of visible matter (see Sects.~\ref{sec:Intro},~\ref{sec:DarkMatter}), although actually the visible matter is converted from (or converts to) the dark matter.  

\newpage
\section{Dark Matter}
\label{sec:DarkMatter}

\paragraph{Visible and invisible (dark) matter.}

According to \emph{the Everett's interpretation} of quantum mechanics, alternative classical realities (Everett's worlds) are ``equally real''. In another wording \cite{MBM2000}, \emph{the classical alternatives objectively coexist}, but they are \emph{separated in consciousness} (in the author's \emph{Extended Everett's Concept}, even more strong assertion is accepted: consciousness \emph{is} the separation of the alternatives, see \cite{MBM2000, MBMbk2010, MBMConsInJOC2011}). This creates subjectively \emph{illusion that only a single alternative exists}. 

Just the same must be valid for \emph{cosmic alternatives}. When we explore astrophysical objects with the aid of our instruments, we observe these objects in the states they have in the subjectively perceived alternative. This alternative (Everett's world) may be called \emph{visible alternative}, and the configuration of matter observed in this alternative, \emph{visible matter}. 

``Other alternatives'' also objectively exist, as well as the configurations of matter in them (remark however that these are not other physical worlds but other classical states of the only existing physical world
, see Sect.~\ref{sec:EverettSQuantumCosmology}). What role these ``other'' (other than one subjectively perceived) cosmic alternatives play in our observations? 

If the Everett's interpretation is applied to quantum cosmology along with the usual way of quantization of gravity, then ``other alternatives'' play no role in our observations. Each of the alternative configurations of matter is accompanied by its own (generated by it) gravitational field. If one of the alternatives (together with the corresponding gravitational field) is subjectively perceived, all the rest are simply unobservable (as well as their own gravitational fields). 

Situation is fundamentally different in the case of the approach we support here. If combining the Everett's interpretation with the semiclassical gravity, we have to conclude that \emph{the matter in ``other alternatives'' cannot be observed with the help of non-gravitational fields} (for example, with the help of light rays or other forms of the electromagnetic field). In this sense \emph{configurations of matter in ``other alternatives'' are invisible}. 

However, since we assume that the common gravity is created by the matter belonging to all alternatives, the invisible matter (the matter in ``other alternatives'') may be ``felt'' by its contribution into the gravitational field. This is why the invisible matter may create the effect known as \emph{the effect of dark matter}. However, our way of explaining this effect requires no special types of matter weakly interacting with the matter of usual type.

\paragraph{Dark halo around visible matter.}

Let us discuss the effect created by invisible matter (the matter in ``other alternatives'') in a more concrete way. Consider some cluster of visible matter (i.e. of matter in the ``visible'' alternative, which is observed by means of the non-gravitational fields). Let us say, in order to be concrete, that this is a galaxy observed in a certain space-time point. What is in the same point (or close to it) in ``other'' alternatives which are invisible for us? 

In some of the other alternatives there is no galaxy in this point. However, in many alternatives the galaxies, of the same size or close to this size, are present in this point or close to this point (we shall see below why the sizes and locations of the galaxies in all these alternatives differ not too strongly). These galaxies are invisible for us (unobserved with the aid of non-gravitational fields). However, their existence in the observed region may be discovered if the gravitational field in this region is measured. Indeed, according to Eq.~(\ref{EinsteinSemiClassic}), this field is created by the matter in all alternatives (not only by the visible matter, i.e. by the matter in ``our'' alternative). 

Gravitational field in the observed region will be created by the mass distribution obtained by averaging over all alternatives. Therefore, this mass distribution includes the contribution from the visible galaxy as well as from those galaxies which belong to other alternatives and are therefore invisible. All these galaxies slightly differ by their sizes and locations. The halo will be approximately spherical (even for non-spherical galaxy) because no space direction is distinguished. 

Therefore, the gravitational field observed in the given region will be such, as if it is created with the visible galaxy and a halo around it consisting of invisible galaxies. This qualitatively corresponds to the observations of \emph{halos of dark matter around visible galaxies}. 

The same is valid not only for galaxies but also for clusters of matter of different scales: \emph{clusters or superclusters of galaxies}. 


\paragraph{Cosmic alternatives become similar in the course of time.}

In the above argument we assumed that the properties of a given galaxy (its size, location etc.) in various alternatives are not too far from each other. This assumption is justified by the fact that the mechanical characteristic of different cosmic alternatives (distributions of mass in them) become closer to each other in the course of time. 

Let us justify this affirmation. We shall do this for a galaxy and its halo. However, the same arguments and conclusions are valid not only for galaxies but for clusters of matter of other scales (galaxy clusters and superclusters).

The key point is that gravitational field is common for all alternatives, and the motion of the matter in each alternative is determined by this common gravitational field. To consider qualitatively the resulting motions, we have to take into account only the inhomogeneity of matter at a certain scale. Let it be a cluster consisting of galaxies belonging to various alternatives but located not far from each other (as compared to the galaxies consisting another cluster). How the parameters of this cluster will change in time?

The matter in any given alternative moves to the center of the cluster consisting of matter of all alternatives. Gradually the galaxies belonging to various alternatives will be shifted closer to this center. The diameter of the cluster will gradually decrease, and the cluster will become more compact. Since galaxies in the cluster belong to various alternatives, decreasing the cluster diameter means that difference between locations of galaxies in different alternatives become smaller. The distributions of matter in different alternatives will gradually become similar. In the limit of very long time, the masses in all alternatives will be distributed in the same way. 

The special role for formating halos is played by \emph{supermassive black holes} in their centers. The crucial circumstance is that a black hole is geometric object rather than material one. It is characterized by an event horizon and a region inside the horizon with such geometric properties that even light cannot leave it. Therefore, a black hole is a special configuration of gravitational field. Being geometrical (gravitational) in its nature, the formation of a supermassive black hole is common the phenomenon for all alternatives. The black hole is formed as a result of collapsing matter from all alternatives. 

At the same time, the black hole acts by its gravitational field on matter in all alternatives. Therefore, the supermassive black hole forms \emph{a sort of anchor} that attaches matter of the visible galaxy and its halo (i.e. matter of all alternatives) to a certain space region (interior region of the black hole) and makes the galaxy and its halo to shrink to this region as a center. 

\newpage
\section{Conclusion}
\label{sec:Conclusion}

Many astrophysical observations confirm what is known now as the effect of dark matter. The most popular explanation of the effect is that, besides the well known (``visible'') matter forming stars and galaxies, there exists another type of matter which possesses similar properties in respect to gravity but does not interact (or interacts weakly) with the visible matter in non-gravitational way. Such matter would not be observed with the aid of the instruments working in various diapasons of electromagnetic field and therefore may be called dark matter. Yet this type of matter might be disclosed because of its gravity. For example, dark matter may consist of the particles predicted by quantum field theory but not up to now discovered in experiments at accelerators. 

We saw in the present paper that a qualitatively different explanation may be obtained on the basis of the Everett's interpretation of quantum mechanics. For this aim one has, in addition to the Everett's (``many-worlds'') form of quantum mechanics, accept the hypothesis of common (for all classical alternatives, or Everett's worlds) gravity. More concretely, one has to assume that the gravitational field has semiclassical character at least in astrophysical scales. 

In the resulting theory, dark matter is interpreted as matter of the same type as the visible matter but existing in those alternative classical realities (Everett's worlds) which are not subjectively perceived. ``Other'' (not subjectively perceived) Everett's worlds are invisible (cannot be observed with the aid of non-gravitational fields) but revealed by their gravitational effect. The matter in these Everett's worlds is ``dark'' but gravitating. 

In subsequent papers we shall discuss the possibility to apply the Everett's semiclassical gravity for explaning observable effects other than the phenomenon of dark matter (see for example the remark at the end of Sect.~\ref{sec:SemiclassicalGravity} about possible non-conservation of energy in the visible alternative).  

\newpage

\appendix

\section{Objective localization and Path Group}
\label{sec:PathGroupLocalization}

\paragraph{Observations in a curved universe in terms of Path Group.}

It is well known that coordinates have no objective meaning in theory of gravity because of the space-time being curved. How then can one objectively characterize localization in space-time? This may be made by describing those operations that a local observer performs during his observation. 

This may be illustrated in a simple case of objective description of localization on the surface of the globe. Let a person (having the role of an observer) be located at the intersection of equator with the Greenwich meridian. If his friend pilot says him: ``to arrive here, I flied 1800 km to the west'', then this precisely determines the point on the surface of the globe his friend arrived from. The curvature of the globe does not prevent the precise localization of the initial point of the rout. The localization of the initial point of the travel will be precisely determined also if the pilot says: ``I flied  1500 km to the west, then turn by 30 degrees to the left and flied 2100 km in the new direction, then turn by 60 degrees to the right and flied 1300 km''. 

The plans of any of these two (and any similar) travels may be presented graphically on the flat map. The map being flat does not prevent correct usage of the information about the lengths of the straight sections of the route and the angles of turns. Although the plan of the travel is drawn on the flat space (the list of paper), it precisely determines (in the corresponding scale) the way at the curved space (surface of the earth) provided that the end point of the way and the direction of the final segment of the path are given. Instead, the initial point of the travel and the direction of the initial part of the way may be given to unambiguously determine the whole path. 

Analogously any, even very complicated, rout through a curved space-time, as well as the initial point of this rout, may be objectively described with the aid of a curve in the Minkowski space-time. The latter may be interpreted as a tangent space to the curved space-time at the point where the observer is located. Even if the geometry of the curved space-time is unknown, the point where the observed object is located, is unambiguously presented if the observer specifies in this way the path through which he achieves the observed object by his instruments (including light rays). 

This simple arguments essentially define how an observer interpret his instrumental astrophysical observations. In fact he make use in this interpretation a natural mapping of curves in a flat space onto the curves in the curved space of the same dimension (given the end point of the path and the orthonormal basis of tangent vectors in it). Besides the dimension, geometry of the space may be arbitrary (but of course smooth). The precise definitions of all operations are given in theory of \emph{Path Group} proposed by the author \cite{MenBk83eng, PathGroupMBM2002}. The paths in the Minkowski space-time may be called \emph{``flat models''} of the corresponding curves in the curved space-time. They form a infinite-dimensional group called Path Group. 

\paragraph{Localization by paths in various alternatives with common gravity.}

We considered above the situation where an observer, located somewhere in the space-time, explores a distant region of this space-time with the help of instruments. The action of the instruments may be presented then by some curve in Minkowski space because this curve unambiguously determines the corresponding curve in the curved space-time (in the physical world) as well as the observed region where this curves starts. In this case we say about real curve and its flat model. 

This is valid, i.e. mapping of flat models onto real curves is unambiguous, provided that 1)~geometry of the physical world, 2)~location of the observer, and 3)~the orthonormal basis of the tangent space in this point are fixed. In this case flat models unambiguously determine the corresponding real curves. Since the end point of the real curve (the location of the observer) is fixed, the flat model unambiguously determines also the location of the region which is observed. The flat model may be interpreted then as an objective characterization of the observed region location. 

If the flat model of a curve is given, but the geometry of the physical world is unknown, then the initial point of the real curve may differ depending on the geometry (gravitational field). This is very special situation when the objective characterization of the actions undertaken by the observer are given, but observed region location remains unknown. 

Let us apply this consideration to the concept of cosmic alternatives. If (as it is assumed in conventional approaches) gravitational fields are different in different alternatives, then the fixation of the flat model (the graphical account of the observer's procedure) cannot be used as an objective characteristic of the localization of the observed object. The flat model determines the observed point only if a certain alternative (and therefore the corresponding geometry) is give in addition to the flat model. This is a subjective characteristic. the flat model does not give purely objective characterization of the observation location. 

the situation is quite different if we accept the approach described in the present paper, i.e. assume not only Everett's interpretation of quantum mechanics but also semiclassical gravity (see Sect.~\ref{sec:EverettSQuantumCosmology}). In this case gravitational field is common for all alternatives. Therefore, fixation of the flat path (Minkowskian curve) is an objective characterization of the observed region location, i.e. the distant point in our physical world (Alterverse). 

This allows one to speak of various alternatives in the same space-time region of our world or about different localizations of a galaxy as it is presented in different alternatives. A galaxy in the subjectively perceived alternative is visible, but the same galaxy as it is presented in another alternative is an element of dark matter. The images of a single galaxy in various alternatives may be a little bit displaced in respect to each other. In result, the dark halo of a visible galaxy is larger than the galaxy itself. We argued in Sect.~\ref{sec:DarkMatter}, that any halo shrinks with time, tending to the size of a single galaxy.

\newpage


\end{document}